\documentclass[twocolumn,epjc3,a4paper]{svjour3}
\smartqed  
\RequirePackage{graphicx}
\usepackage{amsmath}
\usepackage{amsfonts}
\usepackage{amssymb}
\usepackage{xcolor}
\RequirePackage{latexsym}
\RequirePackage[numbers,sort&compress]{natbib}
\RequirePackage[colorlinks,citecolor=blue,urlcolor=blue,linkcolor=blue]{hyperref}
\newcommand\keVee{\mathrm{keV}_\mathrm{ee}}
\newcommand\keVnr{\mathrm{keV}_\mathrm{nr}}
\newcommand\gev{\mathrm{GeV/}c^2}
\journalname{Eur. Phys. J. C}
\begin{document}

\title{Improved EDELWEISS-III sensitivity for low-mass WIMPs using a profile likelihood approach}

\author{
		L. Hehn\thanksref{e1,addr1}
		\and
		E.~Armengaud\thanksref{addr2}
		\and
		Q.~Arnaud\thanksref{addr3,addr13}
		\and
		C.~Augier\thanksref{addr3}
		\and
		A.~Beno\^{i}t\thanksref{addr4}
		\and
		L.~Berg\'{e}\thanksref{addr5}
		\and
		J.~Billard\thanksref{addr3}
		\and
		J.~Bl\"{u}mer\thanksref{addr6,addr1}
		\and
		T. de~Boissi\`{e}re\thanksref{addr2}
		\and
		A.~Broniatowski\thanksref{addr5,addr6}
		\and
		P.~Camus\thanksref{addr4}
		\and
		A.~Cazes\thanksref{addr3}
		\and
		M.~Chapellier\thanksref{addr5}
		\and
		F.~Charlieux\thanksref{addr3}
		\and
		M.~De~J\'{e}sus\thanksref{addr3}
		\and
		L.~Dumoulin\thanksref{addr5}
		\and
		K.~Eitel\thanksref{addr1}
		\and
		N.~Foerster\thanksref{addr6}
		\and
		J.~Gascon\thanksref{addr3}
		\and
		A.~Giuliani\thanksref{addr5}
		\and
		M.~Gros\thanksref{addr2}
		\and
		G.~Heuermann\thanksref{addr6}
		\and
		Y.~Jin\thanksref{addr7}
		\and
		A.~Juillard\thanksref{addr3}
		\and
		C.~K\'{e}f\'{e}lian\thanksref{addr3,addr6}
		\and
		M.~Kleifges\thanksref{addr8}
		\and
		V.~Kozlov\thanksref{addr1}
		\and
		H.~Kraus\thanksref{addr9}
		\and
		V.~A.~Kudryavtsev\thanksref{addr10}
		\and
		H.~Le-Sueur\thanksref{addr5}
		\and
		S.~Marnieros\thanksref{addr5}
		\and
		X.-F.~Navick\thanksref{addr2}
		\and
		C.~Nones\thanksref{addr2}
		\and
		E.~Olivieri\thanksref{addr5}
		\and
		P.~Pari\thanksref{addr11}
		\and
		B.~Paul\thanksref{addr2}
		\and
		M.-C.~Piro\thanksref{addr5,addr14}
		\and
		D.~Poda\thanksref{addr5}
		\and
		E.~Queguiner\thanksref{addr3}
		\and
		S.~Rozov\thanksref{addr12}
		\and
		V.~Sanglard\thanksref{addr3}
		\and
		B.~Schmidt\thanksref{addr1,addr15}
		\and
		S.~Scorza\thanksref{addr6}
		\and
		B.~Siebenborn\thanksref{addr1}
		\and
		D.~Tcherniakhovski\thanksref{addr8}
		\and
		L.~Vagneron\thanksref{addr3}
		\and
		M.~Weber\thanksref{addr8}
		\and
		E.~Yakushev\thanksref{addr12}
}

\thankstext{e1}{e-mail: lukas.hehn@kit.edu}

\institute{Karlsruher Institut f\"{u}r Technologie, Institut f\"{u}r Kernphysik, Postfach 3640, 76021 Karlsruhe, Germany \label{addr1}
			\and
			CEA Saclay, DSM/IRFU, 91191 Gif-sur-Yvette Cedex, France \label{addr2}
			\and
			Univ Lyon, Universit\'{e} Claude Bernard Lyon 1, CNRS/IN2P3, Institut de Physique nucl\'{e}aire de Lyon, 4 rue Enrico Fermi, F-69622 France \label{addr3}
			\and
			Institut N\'{e}el, CNRS/UJF, 25 rue des Martyrs, BP 166, 38042 Grenoble, France \label{addr4}
			\and
			CSNSM, Univ. Paris-Sud, CNRS/IN2P3, Universit\'{e} Paris-Saclay, 91405 Orsay, France \label{addr5}
			\and
			Karlsruher Institut f\"{u}r Technologie, Institut f\"{u}r Experimentelle Kernphysik, Gaedestr. 1, 76128 Karlsruhe, Germany \label{addr6}
			\and
			Laboratoire de Photonique et de Nanostructures, CNRS, Route de Nozay, 91460 Marcoussis, France \label{addr7}
			\and			
			Karlsruher Institut f\"{u}r Technologie, Institut f\"{u}r Prozessdatenverarbeitung und Elektronik, Postfach 3640, 76021 Karlsruhe, Germany \label{addr8}
			\and
			University of Oxford, Department of Physics, Keble Road, Oxford, OX1 3RH, UK \label{addr9}
			\and
			University of Sheffield, Department of Physics and Astronomy, Hicks Building, Hounsfield Road, Sheffield, S3 7RH, UK \label{addr10}
			\and
			CEA Saclay, DSM/IRAMIS, 91191 Gif-sur-Yvette Cedex, France \label{addr11}
			\and			
			JINR, Laboratory of Nuclear Problems, Joliot-Curie 6, 141980 Dubna, Moscow Region, Russian Federation \label{addr12}
			\and
			Now at Queen's University, Kingston, Canada\label{addr13}
	        \and
			Now at Rensselaer Polytechnic Institute, Troy, NY, USA\label{addr14}
	        \and
	        Now at Lawrence Berkeley National Laboratory, Berkeley, CA, USA\label{addr15}
}

\date{September 20, 2016}

\maketitle

\begin{abstract}
We report on a dark matter search for a Weakly Interacting Massive Particle (WIMP) in the mass range $m_{\chi} \in [4, 30]\,\gev$ with the EDELWEISS-III experiment. A 2D profile likelihood analysis is performed on data from eight selected detectors with the lowest energy thresholds leading to a combined fiducial exposure of 496\,kg-days. External backgrounds from $\gamma$- and $\beta$-radiation, recoils from $^{206}\mathrm{Pb}$ and neutrons as well as detector intrinsic backgrounds were modelled from data outside the region of interest and constrained in the analysis. The basic data selection and most of the background models are the same as those used in a previously published analysis based on Boosted Decision Trees (BDT)~\cite{Armengaud:2016cvl}. For the likelihood approach applied in the analysis presented here, a larger signal efficiency and a subtraction of the expected background lead to a higher sensitivity, especially for the lowest WIMP masses probed. No statistically significant signal was found and upper limits on the spin-independent WIMP-nucleon scattering cross section can be set with a hypothesis test based on the profile likelihood test statistics. The 90\% C.L. exclusion limit set for WIMPs with $m_\chi = 4\,\gev$ is $1.6 \times 10^{-39}\,\mathrm{cm^2}$, which is an improvement of a factor of seven with respect to the BDT-based analysis. For WIMP masses above $15\,\gev$ the exclusion limits found with both analyses are in good agreement.
\end{abstract}

\section{Introduction}
\label{sec:introduction}
Through different astrophysical observations on a wide range of cosmological scales, it is well established that $\sim 27\%$ of the energy density in the Universe is made up of an unknown dark matter~\cite{Adam:2015rua}. A well-motivated class of particles proposed to solve the dark matter problem are \textit{Weakly Interacting Massive Particles} (WIMPs) with masses of the order of $\mathrm{GeV}/c^2$ to $\mathrm{TeV}/c^2$ and an extremely low scattering cross section with ordinary matter. Direct detection experiments search for the elastic scattering of a WIMP from the galactic dark matter halo in detectors on Earth-based experiments. The nuclear recoils from such interactions would have an exponentially falling energy spectrum up to a few keV, depending on the mass $m_\chi$ of WIMPs. In addition, the expected rate is smaller than one interaction per kg of target material per year. To minimize the background for this rare event search, the EDELWEISS experiment is located in the Modane Underground Laboratory (LSM) in the French-Italian Alps, where a rock overburden of $4800\,\mathrm{m\ w.e.}$ reduces the cosmic muon flux down to $5\,\rm{muons/m^2/day}$. Remaining muons are tagged with an active muon veto system surrounding the experiment~\cite{Schmidt:2013gdc}, followed by $50\,\mathrm{cm}$ of polyethylene and $20\,\mathrm{cm}$ of lead to suppress neutrons and gammas. Inside these layers of shielding a cryostat made of ultra-pure copper houses germanium monocrystals which are cooled down to $18\,\mathrm{mK}$. A simultaneous measurement of the heat and ionization energies produced in a recoil allows to discriminate between the dominant \textit{electron recoils} (ER) from radioactivity and \textit{nuclear recoils} (NR), which at low energies are only caused by neutrons and the expected WIMP signal.\newline
Other direct detection dark matter experiments use similar approaches based on the same principle to discriminate between different backgrounds and a possible signal from WIMPs. Exclusion limits on the WIMP-nucleon spin-independent scattering cross section from LUX~\cite{Akerib:2015rjg} and SuperCDMS~\cite{Agnese:2014aze} are in strong tension with favoured parameter regions based on observations by DAMA/LIBRA~\cite{Bernabei:2013xsa}, CoGeNT~\cite{Aalseth:2014eft} and CDMSII-Si~\cite{Agnese:2013rvf}.\newline 
Almost all existing signal claims for low-mass WIMPs can be excluded at 90\% C.L. with the improved limits that were recently published by the EDELWEISS-III collaboration~\cite{Armengaud:2016cvl} considering standard assumptions about the WIMP-nucleus interaction and the galactic halo model. Data from a 10-month WIMP search run were analysed in terms of low-mass WIMPs with masses $m_\chi \in [4, 30] \, \gev$ using a method based on Boosted Decision Trees (BDT). No statistically significant excess of events was observed for eight selected detectors, resulting in exclusion limits up to a factor 40 stronger at $m_\chi = 7 \, \gev$, compared to results from previous EDELWEISS-II~\cite{Armengaud:2012pfa} low-energy data. Such a cut-based analysis performs well when the separation of signal and background is sufficient, as is the case for higher WIMP masses. However, at low energy, the finite resolutions of the detectors cause the electron and nuclear recoils to have overlapping populations in the distributions of the variables that serve as discriminator. A separaration thus requires a cut at lower energy, resulting in a severely reduced efficiency. To overcome this problem, the analysis presented here uses an alternative approach which is based on the maximum likelihood, similar to e.g.~\cite{Akerib:2015rjg, Aprile:2011hx}. It is an unblind analysis performed on a similar data sample that was recorded with the same detectors as in~\cite{Armengaud:2016cvl}. With its completely different analysis approach it improves the sensitivity for low-mass WIMPs and allows to cross-check the results of the BDT-based analysis. Instead of extracting limits without background subtraction from a smaller signal region with optimized signal-to-background ratio, the maximum likelihood method is used to model and fit the data in the entire region of interest (RoI). Thus, the remaining WIMP signal after detector efficiency corrections is not further reduced, while expected backgrounds are fitted and can be subtracted. The systematic uncertainties of the background predictions are taken into account by constraints in the likelihood fit and the calculation of exclusion limits.\newline
The operating principle of the EDELWEISS-III detectors and the selection criteria for the analysed data are detailed in Sec.~\ref{sec:detectors}, while a description of the different background components is presented in Sec.~\ref{sec:models}. The formalism of the likelihood model for the analysis is explained in Sec.~\ref{sec:likelihood}, both for fitting the data to individual detectors, as well as for a combined fit of a common signal to all detectors. We also detail how the exclusion limit is set using a hypothesis test based on the profile likelihood test statistics. A discussion of the fit results and a comparison with the result achieved with the BDT method follows in Sec.~\ref{sec:results}.

\begin{figure}
	\includegraphics[width=1\linewidth]{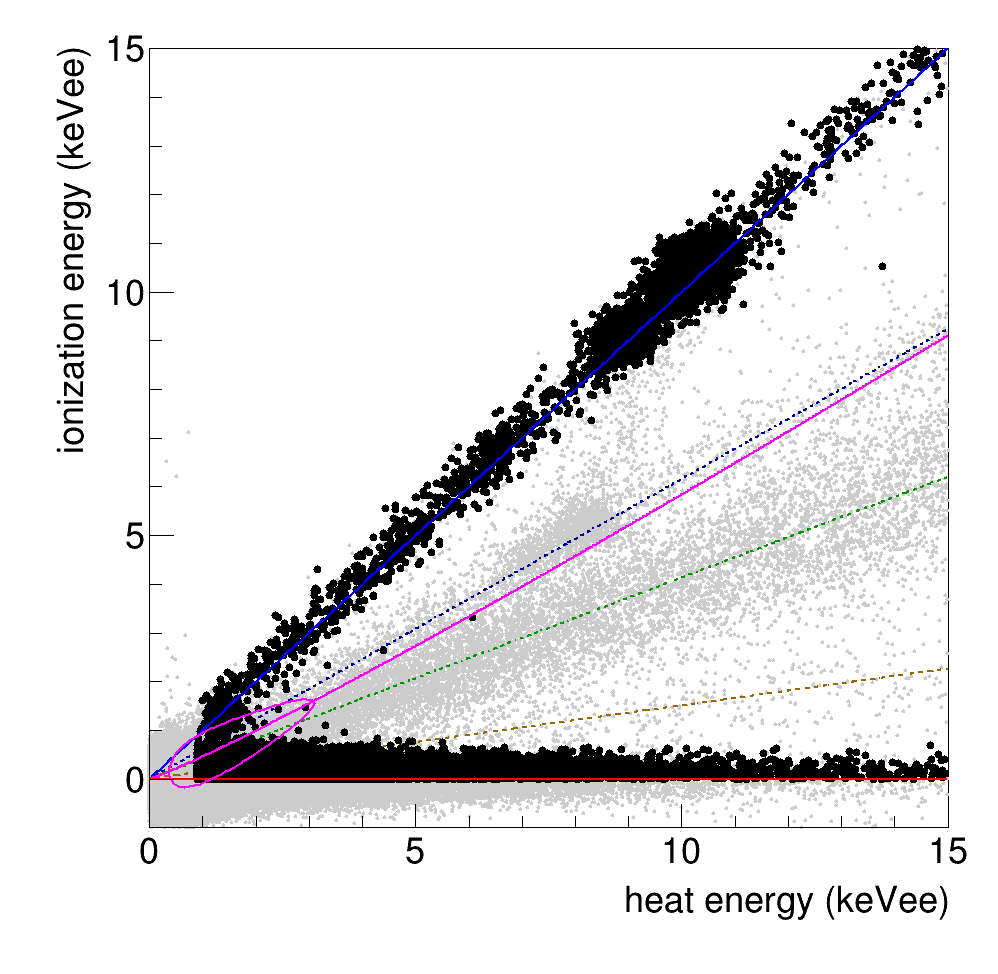} 
	\caption{WIMP search data in the RoI accumulated in eight selected detectors with a fiducial exposure of $496\,\textrm{kg-days}$ in ionization vs. heat energy (black markers). Events before the fiducial cut and in the extended energy range are shown as gray points. Coloured lines indicate the detector-averaged positions that are expected for different background components depending on their ionization yields and collection voltage biases (see text). From top to bottom: electron recoils from tritium decay as well as Compton and cosmogenic gammas in the fiducial volume (blue), surface gammas (dashed blue), nuclear recoils from neutron scattering (magenta), surface betas (dashed green) and $^{206}\text{Pb}$-recoils (dashed brown). Heat-only events have only noise on the ionization channels and no ionization signal on average (red). The coloured contour indicates an $m_\chi = 10\,\gev$ WIMP signal.}
	\label{fig:alldata}
\end{figure}

\section{EDELWEISS-III Detectors and Selection of Data}
\label{sec:detectors}
The detectors used in EDELWEISS-III are of the \textit{Full Inter-Digit} FID800 type~\cite{2012JLTP..167.1056J}. These are high-purity Germanium bolometers in a cylindrical shape of $7\,\text{cm}$ diameter and $4\,\text{cm}$ height with masses ranging from 820 to $890\,\mathrm{g}$ due to small variations both diameter and height. Aluminium electrodes cover all sides of the detector in concentric rings to collect the charge carriers produced in a particle recoil. Glued on the top and bottom surfaces each is a \textit{Neutron Transmutation Doped} (NTD) Ge-sensor, which measures the micro-Kelvin temperature increase due to the energy deposit. The measurement of both heat and ionization signals on an event-by-event basis allows to discriminate rare NR events from the dominating ER events. The latter are mostly due to radioactive background outside the detector which produces $\gamma$-radiation. Their ionization yield $Q$, i.e.\ the ratio of ionization over total recoil energy, is defined as $Q_\mathrm{ER} = 1$ by the energy calibration with a $^{133}\text{Ba}$ $\gamma$-source. The ionization yield $Q_\mathrm{NR}$ for nuclear recoils from neutrons and WIMPs is quenched and thus $\sim 3$ times smaller. It depends on the recoil energy $E_\mathrm{r}$ and can be parameterized as $Q_\mathrm{NR}(E_\mathrm{r}) = 0.16 \cdot \left(E_\mathrm{r}/\mathrm{keV}\right)^{0.18}$ for EDELWEISS Ge-detectors, which is consistent with the Lindhard theory~\cite{Lindhard1963a}. The separation of particle types due to their different $Q$-value is only applicable if the produced charges are properly collected. For recoils close to the detector surface, where charge trapping is important, this is not guaranteed. Therefore, FID detectors are designed to discriminate between surface events and events originating in the bulk of the detector. Interleaved electrode rings on the detector surface are wired together and the resulting groups on the top half are biased with $+4\,\mathrm{V}$, $-1.5\,\mathrm{V}$ and $-4\,\mathrm{V}$, $+1.5\,\mathrm{V}$ on the bottom half. The electric field configuration divides the detector into a \textit{fiducial} volume from which charges are drifted to the so-called \textit{fiducial electrodes} (at $\pm4\,\mathrm{V}$) and a near-surface volume, for which a signal is also seen on the \textit{veto electrodes} (at $\mp1.5\,\mathrm{V}$). This fiducialization is used to significantly suppress backgrounds from surface interactions and select only events with full charge collection efficiency. In this analysis, we consider the heat signal as the resolution weighted average heat energy of the two NTDs and the ionization signal as the averaged fiducial ionization energy of events coming from the fiducial electrodes. These two observables, named $E_\mathrm{heat}$ and $E_\mathrm{ion}$ in the following, are both in units of $\keVee$, as they have been calibrated to fiducial electron recoils from a $^{133}\mathrm{Ba}$ $\gamma$-source. A signal on one of the two veto electrodes, i.e.\ $E_\mathrm{veto}$ is only used to reject surface events.\newline

The data analysed in this work was taken between July 2014 and April 2015, when 24 FID800 detectors with full read-out were installed in the cryostat. WIMP search data was recorded for a total of 161 live days, together with an additional 25 days of calibration data from $^{133}\text{Ba}$ $\gamma$- and AmBe neutron-sources. As we are searching for WIMP signals at very low energies, a proper understanding of the trigger for such events is crucial. Out of the 24 installed detectors, 8 were selected~\cite{Armengaud:2016cvl} because of their low trigger thresholds and good noise conditions. Only hourly periods satisfying requirements on the FWHM baseline resolution of the heat and ionization channels ($\mathrm{FWHM}_\mathrm{heat} < 1\,\keVee$ and $\mathrm{FWHM}_\mathrm{ion} < 0.7\,\keVee$) are considered, leading to a live-time ranging from 95 to 133\,days per detector. More details on the trigger condition and the applied quality cuts, as well as the corresponding efficiencies, are given in~\cite{Armengaud:2016cvl}. The RoI for WIMP search is chosen such that the recoil spectrum for WIMPs with masses up to $m_\chi \approx 30\,\gev$ is included with good efficiency of up to 60\%. We define the RoI by requiring the ionization energy to be $0 \leq E_\mathrm{ion} \leq 15\,\keVee$ and the heat energy of events to be below $15\,\keVee$. The analysis threshold in heat energy $E_\textrm{heat}^\textrm{min}$ varies from detector to detector and depends on the efficiency of the online trigger. The EDELWEISS DAQ-system triggers events on each of the two heat channels, with a trigger threshold adapted on the scale of a few minutes to the instantaneous noise conditions. To ensure a high signal efficiency while minimizing systematic effects we define the analysis threshold in heat as the corresponding energy, for which the livetime-averaged trigger efficiency for each individual detector is above 80\%. This leads to values ranging from $E_\textrm{heat}^\textrm{min} = 0.9\,\keVee$ for the best detector up to $E_\textrm{heat}^\textrm{min} = 1.5\,\keVee$ for the one with lowest efficiency. 

The last cut applied on the data is the selection of fiducial events, which is of particular importance for this analysis. The two observables considered in the likelihood analysis are the heat and ionization energies. To select unambiguous fiducial events only, we require the signal on each of the two veto electrodes to be within $\pm1.64\,\sigma_\mathrm{veto}$ of the energy-dependent Gaussian noise, where $\sigma_\mathrm{veto}$ increases with increasing fiducial ionization energy $E_\mathrm{ion}$ as described in~\cite{Arnaud:2016avs}. The cut with a total acceptance of 81\% was chosen after initial tests on a subsample of data and combines a strong discrimination of surface events with an acceptable signal efficiency loss compared to the BDT analysis~\cite{Armengaud:2016cvl}. The fiducial efficiency was determined for each detector from the homogeneously distributed decay of cosmogenically activated isotopes in the crystal: the number of K-shell electron capture (EC) events giving a peak triplet at $10\,\mathrm{keV}$ was fitted with and without fiducial cut. The resulting effective\footnote{The masses are labeled as effective because their values are reduced by the efficiency of the fiducial cut.} fiducial masses found with this method vary between 508 and $562\,\mathrm{g}$ for the eight detectors. The data after all cuts on data quality and noise periods, before and after the application of the fiducial cut is shown in Fig.~\ref{fig:alldata}. The total fiducial exposure is $496\,\textrm{kg-days}$. Different event populations can be observed and, at higher energies, are distinguishable via their ionization yield $Q$. Surface events from $\gamma$- and $\beta$-radiation as well as $^{206}\text{Pb}$-recoils can be efficiently rejected with the fiducial cut. Remaining events between the electron recoils and heat-only populations are natural WIMP candidates. A likelihood analysis can assess the probability of these events to be of signal or background origin. As input to the likelihood model a \textit{probability density function} (PDF) describing each different component is required.

\begin{figure}
	\includegraphics[width=1\linewidth]{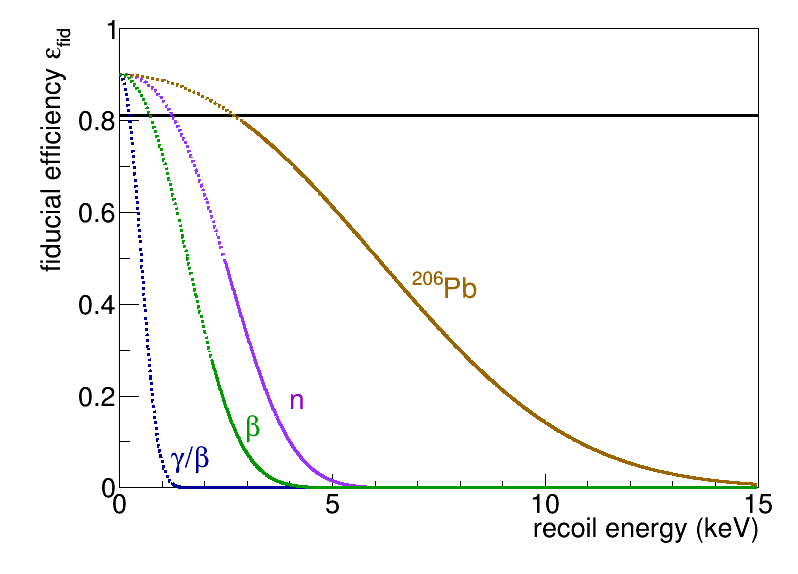} 
	\caption{Survival probability for different surface background components on the top side of detector FID824 as a function of initial recoil energy: electron recoils (blue) from Compton and cosmogenic $\gamma$'s as well as $\beta$'s from the tritium beta decay, nuclear recoils from neutrons or WIMPs (violet), $\beta$'s (green) and $^{206}\mathrm{Pb}$-recoils (brown) from external radioactivity. The dashed part of the efficiency curves is below the analysis threshold $E_\mathrm{heat}^\mathrm{min} = 0.9\,\keVee$ in heat energy (see Eq.~\ref{eq:heatenergy}). Only surface events from $\beta$'s and $^{206}\mathrm{Pb}$-recoils are considered in the analysis as they have a significant contribution after all cuts. For bulk events, the survival probability (or efficiency for the WIMP signal) is considered as approximately constant (black line), neglecting the leakage of surface WIMPs into the acceptance.}
	\label{fig:fiducialefficiency}
\end{figure}

\section{Modelling of the Signal and Background Components}
\label{sec:models}
Thanks to the effective shielding scheme of the EDELWEISS experiment, the residual background mainly originates from radioactive materials inside the cryostat such as connectors, holding structure and detector copper casings as well as from decays of cosmogenically activated isotopes within the detectors~\cite{Armengaud:2013vci,Scorza:2015vla}. Each background component is modelled with a data-driven approach: unblinded data from outside the region of interest (sideband data), acquired in the same WIMP run, are fitted and extrapolated to the low-energy RoI considered in the analysis. In order to construct a likelihood model describing the data for each of the eight detectors, a PDF is calculated for each different background component $i$. This PDF $\mathcal P_\mathrm{i}$ describes a recoil spectrum $\rho_\mathrm{i}(E_\mathrm{r})$ in the two observables heat and ionization energy. It takes into account the ionization yield $Q_\mathrm{i}$ for each background, the efficiency of the trigger on the heat channel $\varepsilon(E_\mathrm{heat})$ and the efficiency of the fiducial cut $\varepsilon^\mathrm{fid}(E_\mathrm{r})$ as well as a Gaussian smearing due to the degraded, energy-dependent resolutions $\sigma_\mathrm{heat}$ and $\sigma_\mathrm{ion}$ of a given detector. In the energy range covered by this analysis, the intrinsic widths of the $Q_\mathrm{i}$-distributions of the different populations are small compared to the effect of $\sigma_\mathrm{heat}$ and $\sigma_\mathrm{ion}$, and are neglected. Before normalization, the PDF can be written as:

\begin{equation}
   \begin{split}
	\mathcal P_\mathrm{i} (E_\mathrm{heat}, E_\mathrm{ion}) =
	\frac{ \varepsilon(E_\mathrm{heat}) }{ 2 \pi \sigma_\mathrm{heat} \sigma_\mathrm{ion} }
	\int_{0}^{\infty} \mathrm{d}E_\mathrm{r} \, \rho_\mathrm{i}(E_\mathrm{r}) \, \varepsilon_\mathrm{i}^\mathrm{fid}(E_\mathrm{r}) \\
	\times \exp \left[ 
	- \frac{ (E_\mathrm{heat}- f_\mathrm{i}(E_\mathrm{r}) )^2 }{2 \sigma_\mathrm{heat}^2} 	
	- \frac{ (E_\mathrm{ion}-Q_\mathrm{i} \cdot E_\mathrm{r} )^2 }{2 \sigma_\mathrm{ion}^2}
	\right]
   \end{split}
   \label{eq:pdfdescription}
\end{equation}

\noindent where the function $f_\mathrm{i}(E_\mathrm{r})$ allows to calculate the observed heat signal of a given recoil energy. It includes the additional heating via the Neganov-Luke effect~\cite{Neganov1985,Luke1988a}, produced by the scattering of charges which are collected by electrodes with a differential voltage $U$ (in volts):

\begin{equation}
	f_\mathrm{i}(E_\mathrm{r}) = \frac{ 1 + Q_\mathrm{i}(E_\mathrm{r}) \frac{U_\mathrm{i}}{3} }{1 + \frac{U_\mathrm{ref}}{3}} \cdot E_\mathrm{r}
   \label{eq:heatenergy}
\end{equation}

The selected detectors have an electric potential of $U_\mathrm{ref} = 8\,\mathrm{V}$ between the fiducial electrodes and bulk ER-events were used to calibrate the energy scale of all heat and fiducial ionization channels. For charges created in the near-surface volume, the Neganov-Luke contribution to the heat energy is smaller, due to the reduced potential of only $5.5\,\mathrm{V}$ between fiducial and veto electrodes. It reduces the measured heat energy for surface events, as can be seen in Fig.~\ref{fig:alldata} for the group of 10\,keV cosmogenic peaks at the  surface which are observed at $E_\mathrm{heat} \approx 7.7\,\keVee$. The measured average value of $Q_\mathrm{i}$ for those events is $0.9$. For surface backgrounds from $\beta$'s and $^{206}\mathrm{Pb}$-recoils as well as so-called heat-only events, the spectrum in heat energy was directly extracted from sideband data. For those components the smearing in heat energy is already included and the PDF before normalization can be directly expressed as:

\begin{equation}
   \label{eq:pdfheatspectrum}
   \begin{split}
	\mathcal P_\mathrm{i} (E_\mathrm{heat}, E_\mathrm{ion}) =
	\frac{ \varepsilon(E_\mathrm{heat}) }{ \sqrt{2 \pi} \sigma_\mathrm{ion} } \,
	\rho_\mathrm{i}(E_\mathrm{heat}) \, \varepsilon_\mathrm{i}^\mathrm{fid} \left(f_\mathrm{i}^{-1}\left(E_\mathrm{heat} \right)\right) \\
	\times \exp \left[	
	- \frac{ (E_\mathrm{ion}-Q_\mathrm{i} \cdot f_\mathrm{i}^{-1}\left(E_\mathrm{heat} \right) )^2 }{2 \sigma_\mathrm{ion}^2}
	\right]
   \end{split}
\end{equation}

\noindent where the average measured value of $Q_\mathrm{i}$ for surface $\beta$'s and $^{206}\mathrm{Pb}$-recoils are 0.4 and 0.1, respectively. The suppression of surface events via the fiducial cut decreases at low energies due to the finite resolution of the ionization channels: the veto energy of a surface event can be smaller than the noise on the veto electrode, thus the event will not be rejected. To a small extent, our data selection is therefore polluted by surface events with heat energies just above the analysis threshold. To build the PDF for these events we take into consideration the efficiency $\varepsilon^\mathrm{fid} (E_\mathrm{r})$ as a function of recoil energy. For surface events the survival probability after the fiducial cut is highly reduced, as is shown in Fig.~\ref{fig:fiducialefficiency} for different background components. It is calculated for each of the detector sides considering the baseline resolution $\sigma_\mathrm{veto}$ and the measured energy $E_\mathrm{veto}$ of the corresponding veto electrode:

\begin{equation}
   \begin{split}
	\varepsilon^\mathrm{fid}_\mathrm{surf} (E_\mathrm{r}) = 
	\frac{1}{ \sqrt{2 \pi} \sigma_\mathrm{veto} } \int_{-1.64 \sigma_\mathrm{veto}}^{+1.64 \sigma_\mathrm{veto}} \text{d}E_\mathrm{veto} \\
	\times \exp \left[ - \frac{(E_\mathrm{veto} - Q_\mathrm{i} E_\mathrm{r})^2}{2 \sigma_\mathrm{veto}^2} \right ] 
   \end{split}
   \label{eq:fiducialcut}
\end{equation}

For events originating in the bulk of the crystal, no signal is measured on the veto electrodes and only noise is reconstructed. The efficiency of the fiducial cut is $\varepsilon^\mathrm{fid}_\mathrm{bulk} = 81\%$ as described in Sec.~\ref{sec:detectors}. The fraction of surface
nuclear recoils leaking into the acceptance below $5\,\mathrm {keV}$
(Fig.~\ref{fig:fiducialefficiency}), and increasing further the WIMP efficiency, is neglected. With the definition of the PDF mentioned above, the WIMP signal and the following background components can be fully described.

\subsection{WIMP Signal}
A signal PDF is constructed for each WIMP mass $m_\chi$ independently, using Eq.~\ref{eq:pdfdescription}. The parametrization for the ionization yield $Q_\mathrm{NR}$ for nuclear recoils has been validated to a precision of 5\% using neutron calibration data taken during the same run. In the description of the signal PDF, $Q_\mathrm{NR}$ is a nuisance parameter and constrained with its systematic uncertainty. The recoil spectrum for the scattering of WIMPs on natural germanium with an average of $A = 72.6$ nucleons is calculated following~\cite{Savage:2007zz}. For all astrophysical parameters we use values corresponding to the Standard Halo Model (SHM), i.e.\ $\rho_\mathrm{DM}^\mathrm{local} = 0.3\,\mathrm{GeV}/\mathrm{c}^2/\mathrm{cm}^3$, $v_\mathrm{0} = 220\,\mathrm{km/s}$, $v_\mathrm{earth} = 230\,\mathrm{km/s}$ and $v_\mathrm{esc} = 544\,\mathrm{km/s}$. With the cuts described in Sec.~\ref{sec:detectors} a potential WIMP signal is reduced to $\sim 60\%$ for $m_\chi = 30\,\gev$. Detector FID824 has the highest sensitivity for a WIMP signal due to its good baseline of the heat channel and the resulting low heat threshold $E_\mathrm{heat}^\mathrm{min} = 0.9\,\keVee$. For this detector, the signal fraction after cuts decreases to $2 \cdot 10^{-4}$ for a $m_\chi = 4\,\gev$ signal but is above 1\% for masses $m_\chi > 5\,\gev$.

\begin{figure}
	\includegraphics[width=1\linewidth]{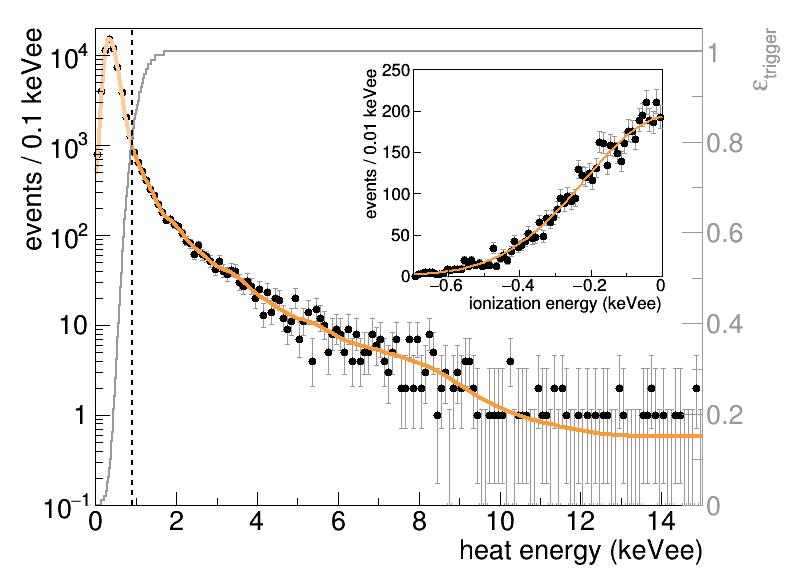} 
	\caption{Heat energy spectrum of events with negative ionization energy (\textit{"heat-only"}) for detector FID824 after quality cuts. Below $E_\mathrm{heat} \approx 1\,\keVee$ the data is dominated by random triggers on noise fluctuations, the efficiency of the heat trigger is given for reference (gray). The spectrum is modelled with a \textit{Kernel Density Estimation} function (orange). Of the 63,400 events in the sideband, 5386 are above the 80\% efficiency analysis threshold of $E_\mathrm{heat}^\mathrm{min} = 0.9\,\keVee$ for this detector (dashed line). Inset: ionization energy spectrum of events above the analysis threshold fitted with a gaussian function of fixed width. The fitted mean of $0.021 \pm 0.005\,\keVee$ results in a systematic error of $14.9\%$ for the expected number of events in the RoI.}
	\label{fig:heatonly_spectrum}
\end{figure}

\begin{figure}
	\includegraphics[width=1\linewidth]{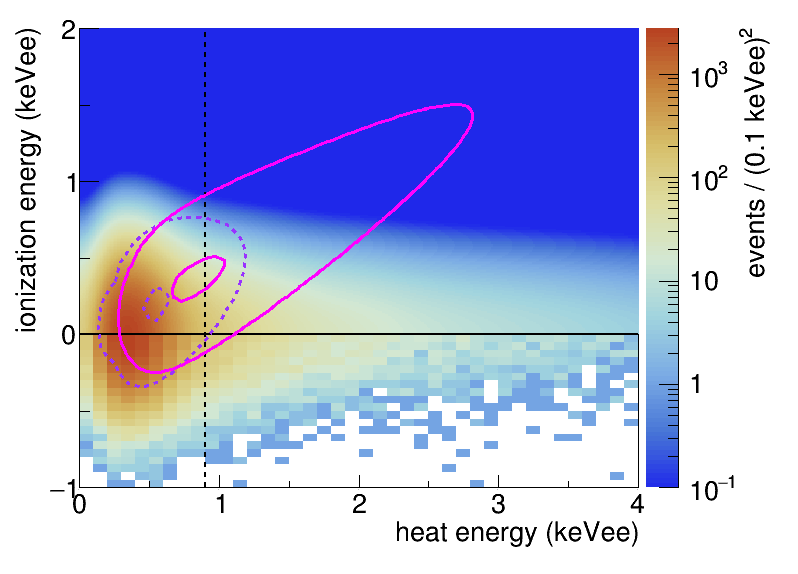} 
	\caption{Sideband data ($E_\mathrm{ion} < 0\,\keVee$, data bins) and modelled heat-only PDF in the RoI ($E_\mathrm{ion} > 0\,\keVee$) for detector FID824. The regions containing 10\%/90\% of the signal density for a WIMP  with $m_{\chi} = 5\,\gev$  and $m_{\chi} = 10\,\gev$ are shown as dashed purple and solid magenta contours, respectively. The dashed black line indicates the analysis threshold in heat energy for this detector. Any signal contribution to the sideband can be considered negligible for the cross sections probed in this analysis.}	 
	\label{fig:heatonly}
\end{figure}

\subsection{Heat-Only Events}
The dominant background in the EDELWEISS-III low-energy data are heat-only events. They are present in all detectors with different intensity and constitute between 85\% and 95\% of the events in the RoI after all cuts. For those events the data acquisition was triggered by a clear signal on one or both NTD heat sensors, while only noise can be seen on each of the four ionization channels, and the signals of the two NTDs are compatible. The heat energy spectrum of those events shows an exponential decrease (e.g.~Fig.~\ref{fig:heatonly_spectrum}) for all detectors and overlaps with randomly triggered noise fluctuations near the heat threshold of a detector. The variation of the heat-only event rate shows a common behaviour for all detectors: a simultaneous burst of the rate which coincides with a period of unstable operating conditions due to the cryogenic system followed by an exponential decay with a time constant of around 20 days which is not compatible with any of the known radioactive isotopes in the setup. A particle origin, e.g.\ from $^{206}\mathrm{Pb}$-recoils absorbed in one of the electrodes and producing no ionization signal, can be excluded due to the high rate and temporal behaviour. Internal radiation within the NTD heat sensors is rejected by a cut requiring a coincident signal in both NTDs described in~\cite{Armengaud:2016cvl}. The source of heat-only events is yet unknown, but possible explanations are the creation of phonons from friction of the detector with the holders, or stress near the NTD gluing spot. Several strategies are pursued to identify the origin of those events and to significantly reduce them in future runs. We use the sideband with negative ionization energy to model heat-only events in the RoI. In the absence of a theory to describe the shape of the heat-only energy spectrum, we use a \textit{Kernel Density Estimation} (KDE) function of the data in this sideband to model this background. The ionization energy spectrum has a gaussian shape with a width given by the average baseline noise for the ionization channels. Fitting the distribution of sideband data in $E_\mathrm{ion}$ with a gaussian indicates a small possible shift of the mean with respect to $E_\mathrm{ion} = 0\,\keVee$. That shift is only statistically significant for some of the detectors and is related to a small fraction of ${<}1\%$ uncorrected cross-talk between heat and fiducial ionization channels. The effect of a possible shift on the number of expected events for this background is taken as a systematic uncertainty and ranges between $0.4\%$ and $14.9\%$ and is considered in the constraint for this background. For the most sensitive detector, FID824, Fig.~\ref{fig:heatonly_spectrum} shows heat and ionization energy spectra of the sideband data with the respective models. In principle, the heat-only sideband can be contaminated by underfluctuations of the ionization energy from low-energy event populations with small ionization yield, such as $^{206}\mathrm{Pb}$-recoils and $\beta $-particles. Considering the low number of expected events for these components ($\mathcal{O}(10)$ events above the analysis threshold per detector) compared to the high rate of heat-only events, the effect on the extracted spectrum is negligible. It was also checked that the number of events for a possible WIMP signal of mass $m_\chi$ in the heat-only sideband is negligible for the cross section excluded in the following. The heat-only sideband data ($E_\mathrm{ion} < 0$) and modelled PDF in the RoI ($E_\mathrm{ion} > 0$) are shown in Fig.~\ref{fig:heatonly}, together with WIMP signals for two different masses.

\subsection{Electron Recoils from Gammas and Betas}
\label{sec:ElectronRecoils}
The energy spectrum of electron recoils in the fiducial volume up to $15\,\keVee$ consists of a set of peaks on top of a continuous component. This component is due to the Compton scattering of gamma rays from external radioactive sources and to betas from the decay of $^{3}\mathrm{H}$ inside the detectors~\cite{Armengaud:2016aoz}. The observed peaks are produced by mono-energetic gammas from electron capture reactions within the crystal and result from the activation of different isotopes due to cosmic rays or neutron calibration. The intensity of these peaks is different for each detector and depends on its age and exposure to cosmic rays before installation underground. In the energy range between $5\,\keVee$ and $7.7\,\keVee$ X-rays from the K-shell EC of the isotopes $^{49}\mathrm{V}$ ($E = 4.97\,\mathrm{keV}$), $^{51}\mathrm{Cr}$ ($5.46\,\mathrm{keV}$), $^{54}\mathrm{Mn}$ ($5.99\,\mathrm{keV}$), $^{55}\mathrm{Fe}$ ($6.54\,\mathrm{keV}$), $^{56,57,58}\mathrm{Co}$ ($7.11\,\mathrm{keV}$) and $^{56}\mathrm{Ni}$ ($7.71\,\mathrm{keV}$) are included in the fit as potential peaks. Around $10\,\keVee$ a triplet of $^{65}\mathrm{Zn}$ ($8.98\,\mathrm{keV}$), $^{68}\mathrm{Ga}$ ($9.66\,\mathrm{keV}$) and $^{68}\mathrm{Ge}$ ($10.37\,\mathrm{keV}$) can be resolved, which has corresponding L-shell peaks at $1.10$, $1.19$ and $1.30\,\mathrm{keV}$ (Fig.~\ref{fig:bestfit}, light blue). While the K-shell peaks are well separated from a WIMP signal in the analysis parameter space, the 3 L-shell peaks can have significant overlap with a signal for the lowest WIMP masses probed. Depending on the analysis threshold $E_\textrm{heat}^\mathrm{min}$ of each detector, the fraction of those peaks in the RoI can vary significantly, from almost full coverage to only a tail of the gaussian peak. With the known L/K-shell ratio of $11\%$~\cite{Bahcall:1963zza} and the calculated peak fraction above threshold we extrapolate the rate of L-shell X-ray events. For this we perform a sideband fit of fiducial events in the electron recoil band $3\,\keVee < E_\mathrm{heat}, E_\mathrm{ion} < 30\,\keVee$ with a separate likelihood model including all K-shell peaks, Compton gammas and tritium $\beta$-events. We find the extrapolated rate of tritium decay for each detector to be in agreement within uncertainties with the rates found in~\cite{Armengaud:2016aoz}. Systematic uncertainties for all ER-components in the RoI are propagated from the errors of this sideband fit and are typically $\mathcal{O}(30\%)$.

\subsection{Unrejected Surface Events}
At higher energies, the fiducial cut allows the rejection of all surface events, as they would induce a clear signal $E_\mathrm{veto}$ on one of the two veto electrodes. For low ionization energies however, the rejection can fail. If the ionization energy of a surface event is low enough, so that $E_\mathrm{veto} < 1.64 \, \sigma_\mathrm{veto}$, the event passes the cut. For particle types with low ionization yield $Q_\mathrm{i}$, the produced ionization energy is smaller, and therefore less charge is collected on the veto electrodes to reject surface events. The surface events in this analysis are mostly $^{206}\mathrm{Pb}$-recoils and $\beta$-particles originating from the $^{238}\mathrm{U}$ decay chain of surrounding materials such as $^{222}\mathrm{Rn}$ daughter isotopes~\cite{Scorza:2015vla}. Those particles have a small penetration depth or even scatter on the crystal surface. Another possible component would originate from the electron recoils described in Sec.~\ref{sec:ElectronRecoils}, which are also produced in the near-surface volume. However, due to their high ionization yield of $Q_\mathrm{ER} \approx 1$, the rejection of these surface events above the heat threshold $E_\mathrm{heat}^\mathrm{min}$ is very efficient: the expected number of events in the RoI after the applied fiducial cut was calculated to be well below $10^{-2}$ for all detectors and these events are therefore not considered in the analysis. For both $\beta$'s and $^{206}\mathrm{Pb}$-recoils, the spectrum in heat energy is extracted from a clear selection of surface events with energies $E_\mathrm{veto} > 5\,\sigma_\mathrm{veto}$ and then extrapolated to the lower heat threshold within the RoI. The ionization yield of the events is fitted from the same sideband data. We do not include any uncertainty on the fitted $Q_i$ as it is negligible with respect to the smearing due to the energy resolutions. Both energy spectra and ionization yield are determined for top and bottom surface of each detector independently.

\subsection{Nuclear Recoils from Neutrons}
Neutron background can mimic a WIMP signal, as neutrons can produce single scatter nuclear recoils with the same ionization yield $Q_\mathrm{NR}$ as WIMPs, according to an exponential energy spectrum. We distinguish between two different sources of neutrons in our detectors: muon-induced and radiogenic neutrons. Simulations showed that in the energy range of this analysis, the number of single scattering neutrons induced by muons is compatible with zero after vetoing~\cite{Kefelian:2016srj}. For radiogenic neutrons coming from radioactivity due to ($\alpha$, n) reactions and spontaneous fissions within the cryostat, Monte Carlo simulations have been performed with all known sources to derive their energy spectrum down to the lowest energies. The spectral shape of the radiogenic neutron background shows little dependence on the exact location of individual sources and can be fitted and parametrized by a double exponential law in the energy range of $2$ to $20\,\keVnr$, calibrated for nuclear recoil interactions. The normalization of the spectrum is derived from data taken with 17 detectors during the same EDELWEISS-III physics run. In the energy range of $10$ to $100\,\keVnr$, nine multiple scattering events are found in the $90\%\,\mathrm{C.L.}$ nuclear recoil band for a fiducial exposure of 1309\,kg-days. This number cannot be reproduced with the simulation of all known sources and hints at an additional neutron source in the experiment. The Monte Carlo simulation however is able to reproduce the measured single-over-multiple-ratio within uncertainties. We derive the normalized neutron spectrum for each detector by weighting it with corresponding exposure in the present data set, as well as the single-over-multiple-ratio of 0.45 from simulations. After all cuts and efficiency corrections, the expected background from single scatter neutrons in the RoI is similar for all detectors and has an average value of $\mu^\mathrm{exp}_\mathrm{neutron} = 0.20 \pm 0.07\,\mathrm{events}$ (Tab.~\ref{tab:backgrounds}). Expected rates for individual detectors have a combined uncertainty of $45\%$ coming from the single-over-multiple ratio uncertainty and the statistical error from the measurement of multiples.

\begin{table}
	\caption{Rate of expected events for different types of backgrounds for detector FID824 and all detectors combined. Event rates for component of the same type have been summed up for demonstration purposes only with propagated systematic errors. During fitting all components are considered as separate PDFs with individual constraints. The background model is clearly dominated by heat-only events.}
	\label{tab:backgrounds}
	\renewcommand{\arraystretch}{1.2}
	\begin{tabular}{l c c}
	\hline\noalign{\smallskip}
	Component & FID824 & Combined \\ 
	\noalign{\smallskip}\hline\noalign{\smallskip}
	Heat-only & $5386\pm804\phantom{0}$ & $44122\pm1356\phantom{0}$ \\
	Cosmogenic $\gamma$'s & $176\pm14\phantom{0}$ & $4358\pm77\phantom{00}$ \\
	Compton $\gamma$'s & $41\pm6\phantom{0}$ & $554\pm26\phantom{0}$ \\
	Tritium $\beta$'s & $43\pm14$ & $624\pm77\phantom{0}$ \\
	Surface $\beta$'s & $8.5\pm2.4$ & $21.0\pm3.6\phantom{0}$ \\
	Surface $^{206}\mathrm{Pb}$-recoils & $6.2\pm0.8$ & $35.5\pm1.6\phantom{0}$ \\
	Neutrons & $0.19\pm0.09$ & $1.60\pm0.72$ \\
	\noalign{\smallskip}\hline\noalign{\smallskip}
	All backgrounds & $5661\pm805\phantom{0}$ & $49655\pm1361\phantom{0}$ \\
	Observed events $N$ & $5685$ & $50715$ \\	
	\noalign{\smallskip}\hline
	\end{tabular}
	\renewcommand{\arraystretch}{1}
\end{table}

With respect to the BDT based analysis~\cite{Armengaud:2016cvl}, most of the background components listed above are identical. Deviations are mainly related to the different fiducial cut and the resulting survival probability of background components. The preselection applied before the BDT analysis accepts more surface beta and gamma events than the present stricter fiducial cuts, leaving the BDT a larger population of these events to optimize its multi-parametric selection. The present fiducial
cut effectively removes most of them. For the same reason we do not include so-called triple events with a signal on both fiducial and one veto electrode. Lastly we intentionally differentiate between bulk events from Compton $\gamma$'s and tritium $\beta$'s as two separate components in the likelihood analysis although their energy spectra are approximately degenerated in the RoI. An overview of the expected event rates summarized for different types of backgrounds is given in Tab.~\ref{tab:backgrounds}. The total background for both detector FID824 and all detectors combined is within $1 - 2\%$ agreement with the observed number of events.

\section{Profile Likelihood Analysis}
\label{sec:likelihood}
With the model of the different background components $i$ as PDFs $\mathcal P_\mathrm{i}(E_\mathrm{heat}, E_\mathrm{ion})$, we can now define for each detector a total PDF $\mathcal P_\mathrm{tot}$ as the sum of all backgrounds and the signal for a given WIMP mass $m_\chi$:

\begin{equation}
	\label{eq:pdf}
	\mathcal P_\mathrm{tot} (\sigma, \vec{\mu} \mid m_\chi) = 
	\frac{1}{\nu} \left[ \mu_\mathrm{\chi} \mathcal P_\mathrm{\chi}(m_\chi) + \sum_\mathrm{i}
	\mu_\mathrm{i} \mathcal P_\mathrm{i} \right]
\end{equation}

\noindent where the combined fitted rate of all components is $\nu = \mu_\chi + \sum_\mathrm{i} \mu_\mathrm{i}$, while the number of observed events is $N$. The rate of WIMP events $\mu_\chi(\sigma)$ is a function of the parameter of interest, the WIMP-nucleon cross section $\sigma$, and is proportional to the integrated signal in the RoI. All other fit parameters are the event rates $\vec{\mu}$ for the different background components $i$. The shape of all PDFs is fixed and given by the detector resolutions of heat and ionization energies, as well as the ionization yield and energy spectra for each component. Each of the background event rates $\mu_\mathrm{i}$ is a nuisance parameter and constrained in the fit with a Gaussian constraint term. The expected rate $\mu_\mathrm{i}^\mathrm{exp}$ for each background is calculated by integrating its unnormalized PDF, while the width $\sigma_i$ of the constraint term is given by the combined statistical and systematic error on this value and varies between all backgrounds and detectors. With this information, we can for each detector construct the extended likelihood function in heat and ionization energies:

\begin{equation}
	\begin{split}
	{\cal L} \left(\sigma, \vec{\mu} \mid m_\chi \right) =
	\prod_\mathrm{n=1}^{N} \mathcal{P}_\mathrm{tot} \left(E_\mathrm{heat}^\mathrm{n}, E_\mathrm{ion}^\mathrm{n} \right) \\
	\times \prod_\mathrm{i} \operatorname{Gauss} \left( \mu_\mathrm{i} \mid \mu_\mathrm{i}^\mathrm{exp}, \sigma_\mathrm{i} \right)
	\times \operatorname{Poisson} \left(N \mid \nu \right)
	\end{split}
	\label{eq:singlelikelihood}
\end{equation}

From these likelihood functions with detector specific PDFs and constraint terms, we can construct a joint likelihood function describing the data for all eight detectors combined:

\begin{equation}
	\label{eq:jointlikelihood}
	{\cal L}_\mathrm{comb} (\sigma, \vec \mu_\mathrm{1}, ...,\vec \mu_\mathrm{8} \mid m_\chi) = \prod_{j=1}^{8} {\cal L}_{\rm{j}} (\sigma, \vec \mu_\mathrm{j} \mid m_\chi)
\end{equation}

Here, each detector has its own, independent background and signal PDFs, as well as nuisance parameters and constraint terms. The only common fit parameter shared by all likelihood terms is the WIMP-nucleon scattering cross section $\sigma$. If the fit for a given WIMP mass results in a \textit{maximum likelihood estimator} (MLE) for the cross section $\hat \sigma$ compatible with zero within errors, we can set an exclusion limit. For this we follow~\cite{Cowan:2010js} and perform a hypothesis test based on the profile likelihood ratio $\lambda(\sigma)$:

\begin{equation}
\label{eq:profilelikelihood}
\lambda (\sigma) = \frac{\cal L (\sigma, {\hat{\vec{\mu}}}^\prime)}
{\cal L (\hat{\sigma}, \hat{\vec{\mu}})}
\end{equation}

\noindent where ${\hat{\vec{\mu}}}^\prime$ are the MLE of the nuisance parameters when maximizing the likelihood for a fixed value of the cross section $\sigma$. The test statistics $q_\sigma$ used to reduce the data to a single value is defined as:

\begin{equation}
\label{eq:teststatistics}
q_\sigma =
\begin{cases}
{-2}\ln \lambda(\sigma) &  \sigma \geq \hat{\sigma} \\ 
0 &  \sigma < \hat{\sigma}
\end{cases}
\end{equation}

The probability distribution functions $f(q_\sigma {\mid} H_\sigma)$ and $f(q_\sigma {\mid} H_0)$ under the signal hypothesis $H_\sigma$ and the back\-ground-only hypothesis $H_0$ are used to find the cross section $\sigma$ for which $H_\sigma$ can be excluded at 90\%\,C.L. while correcting for downward fluctuations of the background following the prescription in~\cite{Read:2002hq}. The parametrisation of the probability distribution functions $f(q_\sigma {\mid} H_\sigma)$ and $f(q_\sigma {\mid} H_0)$ with an approximation as described in~\cite{Cowan:2010js} was found to yield limits with a C.L. less than 90\% for some of the probed WIMP masses. We use Monte Carlo generated datasets to derive all upper limits. Calculations of limits are performed with the \texttt{RooStats}-package~\cite{Moneta:2010pm}, based on the \texttt{RooFit}~\cite{Verkerke:2003ir} framework with which all PDFs and likelihood functions are constructed.

\begin{figure}
	\includegraphics[width=1\linewidth]{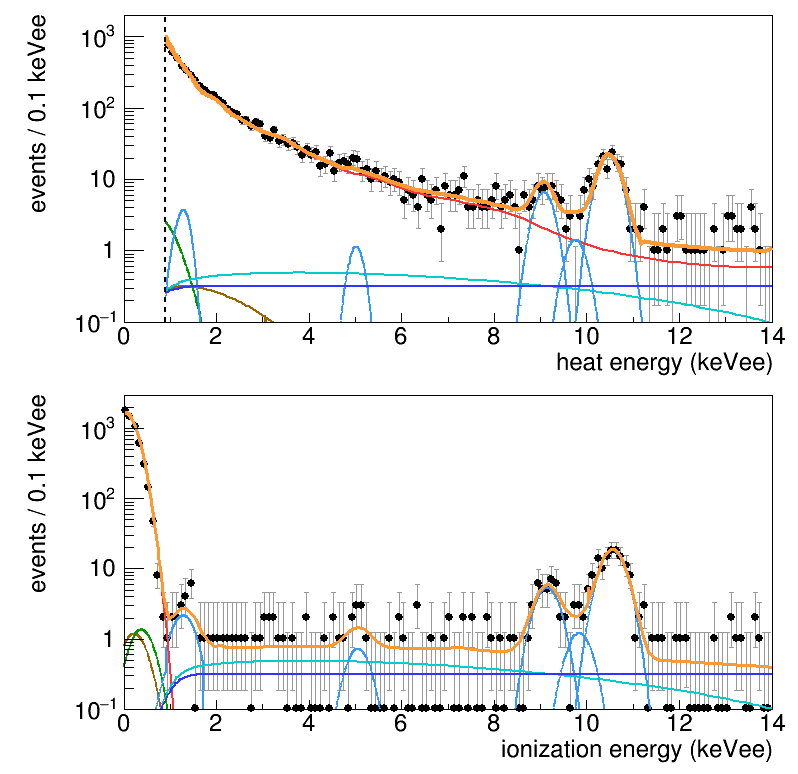} 
	\caption{Energy spectra in the two observables $E_\mathrm{heat}$, $E_\mathrm{ion}$ for single-scattering events in the fiducial volume of detector FID824 passing all quality cuts. The projection of the best fit PDF for an individual fit to this data is shown in orange. The fitted signal component for $m_{\chi} = 4\,\gev$ is zero. All background components are scaled to their corresponding fit values: heat-only (red), Compton gammas (dark blue), tritium $\beta$'s (turquoise), cosmogenic K and combined L-shell peaks (light blue), $\beta$-events (green) and $^{206}\text{Pb}$-recoils (brown). Top: distribution in heat energy, which is dominated by the exponential heat-only spectrum at energies near the analysis threshold (dashed black line). Bottom: distribution in ionization energy showing a clear separation between the Gaussian heat-only noise around $0\,\keVee$ and the electron recoil background.}
	\label{fig:bestfit}
\end{figure}

\begin{table}
	\caption{Analysis threshold $E_\mathrm{heat}^\mathrm{min}$ (in $\keVee$) for all detectors together with the fitted rate $\mu_\chi$ of signal events for the individual fit of data from single detectors and three different WIMP signal masses. For the combined fit over all detectors with a common signal the total signal rate is given together with the corresponding best fit cross section $\sigma$.}
	\label{tab:numbers}
	\renewcommand{\arraystretch}{1.5}
	\begin{tabular}{l r r r r}
	\hline\noalign{\smallskip}
	Detector & $E_\textrm{heat}^\textrm{min}$ & $4\,\gev$ & $10\,\gev$ & $30\,\gev$ \\ 
	\noalign{\smallskip}\hline\noalign{\smallskip}
	FID824 & $0.90$ & $0.0^{{+}4.7}_{{-}0.0}$ & $0.0^{{+}1.4}_{{-}0.0}$ & $0.0^{{+}0.6}_{{-}0.0}$ \\
	FID825 & $1.13$ & $0.0^{{+}3.5}_{{-}0.0}$ & $3.1^{{+}3.3}_{{-}2.1}$ & $1.2^{{+}1.8}_{{-}0.9}$ \\
	FID827 & $1.03$ & $4.2^{{+}22.3}_{{-}\phantom{0}4.2}$ & $0.0^{{+}1.1}_{{-}0.0}$ & $0.0^{{+}0.6}_{{-}0.0}	$ \\
	FID837 & $1.23$ & $0.0^{{+}12.1}_{{-}\phantom{0}0.0}$ & $2.8^{{+}4.2}_{{-}2.8}$ & $2.5^{{+}2.3}_{{-}1.5}$ \\
	FID838 & $1.12$ & $37.6^{{+}27.7}_{{-}26.1}$ & $2.6^{{+}3.5}_{{-}1.9}$ & $2.6^{{+}2.4}_{{-}1.6}$ \\
	FID839 & $1.40$ & $0.0^{{+}16.5}_{{-}\phantom{0}0.0}$ & $0.0^{{+}3.0}_{{-}0.0}$ & $0.0^{{+}0.9}_{{-}0.0}$ \\
	FID841 & $1.19$ & $39.8^{{+}20.5}_{{-}18.8}$ & $1.2^{{+}3.2}_{{-}1.2}$ & $0.0^{{+}0.9}_{{-}0.	0}$ \\
	FID842 & $1.45$ & $6.7^{{+}16.7}_{{-}\phantom{0}6.7}$ & $1.2^{{+}2.6}_{{-}1.2}$ & $0.0^{{+}1.1}_{{-}0.0}$ \\
	\noalign{\smallskip}\hline\noalign{\smallskip}
	\multicolumn{2} {l} {Comb. $\mu_\chi$ (evts)} & $0.0^{{+}15.7}_{{-}\phantom{0}0.0}$ & $9.9^{{+}7.0}_{{-}5.4}$ & $5.3^{{+}3.2}_{{-}3.2}$ \\
	\multicolumn{2} {l} {Comb. $\sigma\,\left(10^{{-}43} \textrm{cm}^2\right)$} & $1.4^{{+}6189}_{{-}1.4}$ & $2.9^{{+}2.0}_{{-}1.6}$ & $0.3^{{+}0.2}_{{-}0.2}$ \\
	\noalign{\smallskip}\hline
	\end{tabular}
	\renewcommand{\arraystretch}{1}
\end{table}

\section{Results}
\label{sec:results}
\begin{figure}
	\includegraphics[width=1\linewidth]{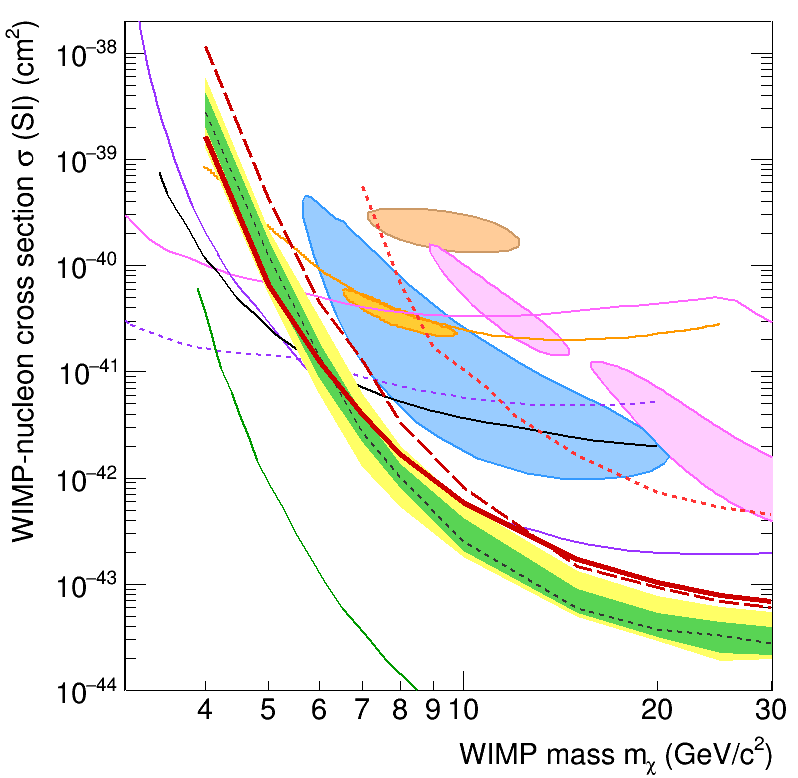} 
	\caption{Calculated $90\% \,\mathrm{C.L.}$ exclusion limit on the spin-independent WIMP-nucleon scattering cross section $\sigma$ as a function of WIMP mass $m_\chi$ for the combined fit over all detectors (solid red). The green and yellow band represent the $1$ and $2\sigma$ confidence band of the expected median sensitivity (dashed black). Shown for comparison is the result of the BDT based analysis~\cite{Armengaud:2016cvl} (dashed red). Contours show possible signals from CDMS-II (Si)~\cite{Agnese:2013rvf} (blue), DAMA~\cite{Savage:2008er} (brown), CRESST-II~\cite{Angloher:2011uu} (pink) and CoGeNT~\cite{Aalseth:2014eft} (orange). Other existing exclusion limits are from EDELWEISS-II~\cite{Armengaud:2012pfa} (small red dashes), CoGeNT~\cite{Aalseth:2012if} (orange), CRESST~\cite{Angloher:2015ewa} (pink), SuperCDMS~\cite{Agnese:2014aze} (purple), XENON100~\cite{Aprile:2016wwo} (black), CDMSlite~\cite{Agnese:2015nto} (dashed violet) and LUX~\cite{Akerib:2015rjg} (green).} 
	\label{fig:exclusion}
\end{figure}

The likelihood fit of the selected data after all cuts is a two-step process: first, an unbinned fit to the data of the eight detectors is performed independently, using the constrained likelihood functions ${\cal L}_\mathrm{j}$ described in Eq.~\ref{eq:singlelikelihood}. It allows to find the best fit values for all of the nuisance parameters describing the different backgrounds for each detector. With these values as starting point, a combined fit over all detectors is performed, in which the individual signal PDFs $\mathcal P_\chi(m_\chi)$ share a common WIMP-nucleon scattering cross section $\sigma$. The result of the individual fit of a WIMP signal with $m_\chi = 4\,\textrm{GeV}$ to data of detector FID824 is shown in Fig.~\ref{fig:bestfit}. Fitted signal rates for three different WIMP signal masses are given in Tab.~\ref{tab:numbers}. For FID824, which is the most sensitive detector due to its low energy threshold $E_\mathrm{heat}^\mathrm{min} = 0.9\,\keVee$, no signal is fitted for any of the probed WIMP masses $m_{\chi} \in [4, 30]\,\rm{GeV}$. The same is valid for detector FID839 for all WIMP masses. For several other detectors (FID\-827, FID838, FID841 and FID842) a strong degeneracy between the WIMP signal and heat-only events is observed. This degeneracy is only present up to WIMP masses of around $10\,\gev$. Therefore, a signal is fitted for these detectors while the rate of heat-only events is underestimated by the same magnitude. However, the resulting signals are always associated with large uncertainties, and they are compatible with zero within $2\,\sigma$. Three detectors (FID825, FID837 and FID838) have events above $E_\textrm{heat} = 2\,\keVee$ which are close to the ionization yield $Q_\mathrm{NR}$ expected for nuclear recoils, as shown in Fig.~\ref{fig:alldata}. For these events a degeneracy between WIMP signal and neutron component is observed, depending on the WIMP mass $m_\chi$. For masses above $m_\chi \approx 10\,\gev$ those events are better described by the signal component, as the expected rate of neutrons is constrained to much lower average values. The resulting excess is between $1 - 3$ signal events which are fitted, depending on the detector and in all cases compatible with no signal. The best fit rates for all other backgrounds are in good agreement with the values expected from their constraint terms, independent of the probed WIMP mass. In the combined fit over all detectors with a common signal cross section $\sigma$, the degeneracies for individual detectors are alleviated and no signal is fitted up to a mass of $m_\chi = 6\,\gev$. For masses above, the aforementioned nuclear recoil candidate events lead to a positive but not significant signal, which peaks at $m_\chi = 7\,\gev$ and then decreases again. Due to the similar exposure and the small influence of the different heat threshold $E_\mathrm{heat}^\mathrm{min}$ for larger mass WIMP signals, the fitted cross section leads to a fairly similar signal rate which is fitted for all eight detectors. Again, all of the fitted signals in the combined fit are compatible with zero.\newline
In the absence of a statistically significant signal for any of the probed WIMP masses, we set 90\%\,C.L. upper limits for the WIMP-nucleon cross section $\sigma$ using the profile likelihood based test statistics described in Eq.~\ref{eq:teststatistics}. The resulting exclusion limit is shown in Fig.~\ref{fig:exclusion}. For masses below $m_\chi = 6\,\gev$ the observed limit is better than the expected median sensitivity due to an underfluctuation of background in the most sensitive detectors. For masses above $m_\chi = 10\,\gev$ the limit is 2 to 3 sigma above the expected sensitivity, due to the presence of NR candidate events in multiple detectors which are in excess with respect to the expected neutron background. This excess is in good agreement with the observations quoted in~\cite{Armengaud:2016cvl}, as both data sets contain these events. However, the 90\%\,C.L. limit of the analysis here presented is a factor of seven stronger for $m_\chi = 4\,\gev$ due to the absence of cuts other than the fiducial selection and the resulting higher signal efficiency as well as a subtraction of the expected backgrounds. For higher masses above $\sim 15\,\gev$ limits from BDT and likelihood approaches are in very good agreement. For these masses the nuclear recoil spectrum of WIMPs extends to high energies, where it can be well discriminated from the dominating heat-only background and the electron recoil component. In this case, the BDT analysis can easily separate a signal with high efficiency and there is no gain in performing a likelihood analysis. The $\sim 15\%$ lower exposure due to the stricter fiducial cut with a total acceptance of $81\%$ is compensated by a higher signal efficiency, and the resulting limits are very similar. We also find a good agreement between the ratios of observed and expected exclusion limits for the two analyses, which we consider to be an inherent property of the data and a confirmation of the validity of this approach.

\section{Conclusion}
\label{sec:conclusion}
We have presented a search for low-mass WIMPs with the EDELWEISS-III experiment, using eight selected detectors and data taken with a total fiducial exposure of 496\,kg-days after all cuts. A data-driven approach was used to model relevant backgrounds from sideband data. For each detector a likelihood function describing the data in heat and fiducial ionization energies was constructed, with constraint terms for each of the nuisance parameters taking into account systematic uncertainties. No statistically significant signal was found, neither for the fit of data from single detectors, nor for a combined fit over all detectors with a common signal cross section. Exclusion limits were set with a hypothesis test using a profile likelihood based test statistics, including corrections for under-fluctuations of the background. At 90\%\,C.L. limit we exclude spin-independent WIMP-nucleon scattering cross sections of $\sigma = 1.6 \times 10^{-39}\,\textrm{cm}^2$ ($6.9 \times 10^{-44}\,\textrm{cm}^2$) for a WIMP mass of $m_\chi = 4\,\textrm{GeV}$ ($m_\chi = 30\,\textrm{GeV}$). Thanks to the higher signal efficiency and a subtraction of the expected backgrounds, the likelihood analysis shows an improvement of a factor of $\sim 7$ for $4\,\gev$ WIMPs compared to a BDT based analysis while reproducing the limit at $15\,\gev$ and above. The results and achieved sensitivity underline the power of a maximum likelihood analysis based on detailed background models.

\begin{acknowledgements}
The help of the technical staff of the Laboratoire Souterrain de Modane and the participant laboratories is gratefully acknowledged. EDELWEISS is supported in part by the German ministry of science and education (BMBF Verbundforschung ATP Proj.-Nr. 05A14VKA), by the Helmholtz Alliance for Astroparticle Physics (HAP), by the French Agence Nationale pour la Recherche (ANR) and the LabEx Lyon Institute of Origins (ANR-10-LABX-0066) of the Universit\'e de Lyon within the program ``Investissements d'Avenir'' (ANR-11-IDEX-00007), by the P2IO LabEx (ANR-10-LABX-0038) in the framework ``Investissements d'Avenir'' (ANR-11-IDEX-0003-01) managed by the ANR (France), by Science and Technology Facilities Council (UK), and the Russian Foundation for Basic Research (grant No. 15-02-03561).
\end{acknowledgements}

\bibliographystyle{h-physrev}
\bibliography{references}   

\end{document}